\documentstyle[preprint,aps]{revtex}

\begin{document}
\title{LONG-RANGE EXCITATION OF COLLECTIVE MODES IN MESOSCOPIC METAL CLUSTERS}
\author{M.S. Hussein}
\address{Instituto de F\'\i sica, Universidade de S\~ao Paulo, \\
C.P. 66318, S\~{a}o Paulo, 05315-970, Brazil}
\author{V. Kharchenko}
\address{Institute for Theoretical Atomic and Molecular Physics\\
at the Harvard-Smithsonian Center for Astrophysics,\\
60 Garden Street, Cambridge, Massachusetts 02136 - U.S.A.}
\author{L.F. Canto and R. Donangelo}
\address{Instituto de F\'\i sica,Universidade Federal do Rio de Janeiro, \\
C.P. 68528, Rio de Janeiro, 21945-970, Brazil}
\date{\today}
\maketitle

\begin{abstract}
We develop a semiclassical theory for the long range excitation of plasmon
resonances in atomic clusters, based on the doorway hypothesis. The effect
of the width of the plasmon resonance is fully taken into account. As an
application we study plasmon excitation in small Sodium clusters, in
collisions with electrons and protons.
\end{abstract}

\section{INTRODUCTION}

Metal atomic clusters have been studied intensively for many years to
explore the difference between atomic-like and bulk material behavior of the
nano-meter metal particles\cite{1}. Intermediate size clusters, being
between molecules and bulk materials, have shown very complex energy spectra
and relaxation dynamics. This complexity results from chaotic behavior of
multi electron energy levels and their interaction with multiple vibrational
modes of the cluster atoms. Experimental data about photo-absorption and
photo-ionization spectra of mesoscopic clusters have exhibited pronounced
structure related to the collective excitation of cluster electrons\cite{1,2}%
. To explain the manifestation of plasmon-like modes in small size atomic
clusters, ``ab initio'' calculations have been done in the framework of the
Functional Density Method (FDM) and the Random Phase Approximation (RPA)\cite
{2}-\cite{4}. Later, a better agreement between theory and experiment has
been reached using Hartree-Fock and RPA evaluations for detailed treatment
of electronic collective modes\cite{5,6}. At the same time, semi-empirical
models have been developed for simple explanations of the experimentally
observed collective phenomena on the base of the surface and volume plasmon
effects\cite{1,7}.

An alternative way to investigate the dynamics of excited cluster states is
through cluster excitation by charged particles: protons, ions, fast
electrons or charged clusters. The involvement of a large number of internal
degrees of freedom influences strongly the cluster response to the external
field. For example, electronic excitation may relax so rapidly among
numerous vibrational modes, that it can change the character of a collective
wave function during an excitation process. We will show, that the dynamics
of cluster excitation can be described as the excitation of some envelope
state of cluster eigenfunctions, even under conditions when the cluster
spectra display a strong chaotic behavior. The scale alteration of the time
and intensity of the cluster excitation by variation of velocity $v$, impact
parameter $b$ and charge $Z$ of projectile particles may be used for
investigations of the dynamics of the cluster excitation and its energy
relaxation. Analysis of the energy for different ($v,Z$) in conjunction with
a description of the structure of the chaotic part of clusters' spectra can
provide the basis for a theoretical description of excitation and decay of
the collective excitations in mesoscopic clusters.

In the present article we have developed a semi-classical method for the
description of the long-range cluster excitation in collisions with charged
particles. Our theoretical approach takes fully into account the finite
lifetime of the excited modes and their reconstruction into chaotic states
of excited clusters. The dynamical description presented in this article may
be considered as concrete realization of the doorway phenomena\cite{8,9}.
Excitation of the doorway mode is well known for inelastic collisions of
nuclear particles as well as for electronic transport through mesoscopic
quantum dots\cite{10}.

The paper is organized as follows: The basic semiclassical reaction theory
is presented in Section II. In Section III we give a full account of the
doorway model of plasmon resonance excitations. This model is then
incorporated, in Section IV, into the reaction theory of Section II, and
application of the theory to several cases are then presented in Section V.
Finally, in Section VI, concluding remarks are made.

\section{PLASMON OSCILLATIONS AS SEMICLASSICAL COULOMB EXCITATION OF ATOMIC
CLUSTERS}

We consider in this section the excitation of collective plasmon resonances
in metal clusters. The probe could be any charged particle (electron,
positron, proton, nucleus, cluster etc.). The theoretical framework of our
development is a time-dependent semiclassical collision theory\cite{AW}. The
collective modes excited in the atomic clusters are treated as damped
vibrations specified by an energy and a decay width. The novel feature of
our theory is the treatment of the life-time of the resonances and
reconstruction of the chaotic part of the cluster in a dynamical way.

We show in figure 1 a schematic representation of our scattering system. The
interaction between the projectile, with charge $Ze$, and the cluster's free
electrons is given by

\begin{equation}
V\left( {\bf r}_{i},{\bf R}\right) =-\sum\limits_{i=1}^{N}\ \frac{Ze^{2}}{%
\left| {\bf R}(t)+{\bf r}_{i}\right| }\ +\ \frac{Ze^{2}}{R(t)}\approx \ 
\frac{Ze^{2}}{R^{2}(t)}\ \sum\limits_{i=1}^{N}{\bf \ \hat{R}}(t)\cdot {\bf r}
_{i}\;,  \label{1}
\end{equation}
where N is the number of free electrons in the cluster, \ ${\bf \hat{R}}(t)$
is the unit vector parallel to the projectile's position vector, ${\bf R}(t)$%
, and ${\bf r}_{i}$ gives the position of the i-th free electron inside the
cluster. Above, the dipole approximation has been introduced. The projectile
motion is described classically and, therefore, once the over-all mean field
potential describing the projectile-cluster interaction is given, ${\bf R}(t)
$ is considered to be known. The interaction, Eq.~(\ref{1}), can be
rewritten as

\begin{equation}
V\left( {\bf r}_{i},{\bf R}\right) =-\,{\bf D\cdot E}(t)\;,  \label{2}
\end{equation}
where ${\bf E}(t)$ is the classical time-dependent electric field vector,

\begin{equation}
{\bf E}(t)=\frac{Ze}{R^{2}(t)}\,\,\,{\bf \hat{R}}(t)  \label{3}
\end{equation}
and ${\bf D}=e\,\sum_{i=1}^{N}\,{\bf r}_{i}\, $ is the quantum mechanical
dipole operator.

Having specified the degrees of freedom of the system, we turn now to the
description of the collision process. The time-dependent Schr\"{o}dinger
equation that describes the evolution of the atomic cluster is

\begin{equation}
\left[H_{0}+V\right] \,\,\Psi ({\bf r}_{1},{\bf r}_{2},..,{\bf r}_{N},\,t)=
i\hbar \,\frac{\partial \Psi ({\bf r}_{1},{\bf r}_{2},..,{\bf r}_{N},\,t)}{%
\partial t},  \label{4}
\end{equation}
where $H_{0}$ is the intrinsic Hamiltonian of the cluster whose spectrum is
given by 
\begin{equation}
H_{0}\vert\varphi _{n}\rangle =\varepsilon_{n}\,\vert\varphi _{n} \rangle \,.
\label{5}
\end{equation}
Making the expansion

\begin{equation}
\left| \Psi \right\rangle =\sum\limits_{n}a_{n}(t)\,e^{-i\varepsilon
_{n}t/\hbar} \mid \varphi _{n}\rangle \,,  \label{6}
\end{equation}
we derive the following familiar set of coupled equations (we are taking $%
\varepsilon _{0}=0$)

\begin{equation}
i\hbar \, \frac{da_{0}(t)}{dt}=\sum\limits_{n}e^{-\,i\varepsilon
_{n}t/\hbar}\; V_{0n}(t)\,\,a_{n}(t)  \label{7}
\end{equation}
\begin{equation}
i\hbar\, \frac{da_{n}(t)}{dt}=e^{\,i\,\varepsilon_{n} t/\hbar}\;
V_{n0}(t)\,\,a_{0}(t)\ +\ \sum\limits_{m\ne 0}e^{i\left(\varepsilon
_{n}-\varepsilon _{m}\right)t/\hbar}\; V_{nm}(t)\,\,a_{m}(t) \\
,.  \label{8}
\end{equation}
These equations are solved with the usual initial conditions

\begin{equation}
a_{n}\left( t\rightarrow -\infty \right) =\delta _{n0} \,,  \label{9}
\end{equation}
and the excitation probabilities are

\begin{equation}
P_{n}=\left| a_{n}(t\rightarrow +\infty )\right| ^{2}\,.  \label{10}
\end{equation}

The expansion in Eqs.~(\ref{7},\ref{8}) is overwhelmingly complicated in the
case of mesoscopic clusters, especially in the region of dynamic excitation.
A convenient way to describe the average behavior of the dynamics is through
the introduction of doorway states, a well known concept in the theory of
nuclear reactions~\cite{8}. In the following, we give an account of this
approach.

\section{DOORWAY STATE DESCRIPTION OF DYNAMICAL COLLECTIVE EXCITATION OF
ATOMIC CLUSTERS}

It is well known that quantal many-body systems exhibit both collective and
statistical, chaotic, behaviors. The interplay between these modes is
clearly exhibited in the response of these systems to external probes. One
sees usually in the spectrum a concentration of dipole oscillation strength
at a given excitation energy which is spread over a rather large interval.
The spread defines the width of the so-called collective excitation which is
connected to the coupling to the more complex, chaotic degrees of freedom.

Both in the theory of nuclear reactions and in the many body theory of atoms~
\cite{4}-\cite{6}, it has been common to give explicit reference to the
collective state, although it is not an eigenstate of the isolated quantal
system. The term `doorway state' is used to label the collective state
usually called Giant Resonance in the nuclear physics context. The only
eigenstates of the systems are the complicated configurations mentioned in
the introduction. However, due to the usually very large number of these
states, it has been proven useful the use of the doorway state to describe
the average excitation of the system. In the following, we develop a
theoretical framework that allows the treatment of the excitation of the
doorway state (collective state) in the collision process. Implicit
reference to the statistical degrees of freedom is allowed through explicit
consideration of the damping width of the doorway.

The starting point of our consideration is Eq.~(\ref{5}), which describes
the spectrum of the atomic cluster. In a collision process, the cluster is
originally in the ground state, $\vert\varphi _{0}\rangle $, and transitions
to excited states take place through the external coupling, $V$. We shall
assume that these excited states are modulated by a doorway, which could be
a plasmon resonance. Further, to reach these `fine structure' states, we
assume that the systems has to go through the doorway(s) $\left| d
\right\rangle $\cite{8,9}. We formally accomplish this by expressing $\left|
\varphi _{n}\right\rangle $ as

\begin{equation}
\left| \varphi _{n}\right\rangle =\alpha _{n}\left| d\right\rangle
+\sum\limits_{m}\beta _{n}^{m}\,\,\left| m\right\rangle \,,  \label{11}
\end{equation}
where the states $\left| m\right\rangle $ form an orthonormal set which
spans the intrinsic subspace orthogonal to $\left| d\right\rangle $. Notice
that neither $\left| d\right\rangle $ nor $\left| m\right\rangle $ are
eigenstates of the system. The doorway nature of $\left| d\right\rangle $ is
accomplished by requiring that $\langle \varphi_0\vert V\vert m\rangle = 0$,
so that the ground state is coupled to the excited states through

\begin{equation}
\langle \varphi _{0}\left| V\right| \varphi _{n}\rangle \equiv V_{0n}=\alpha
_{n}\,\,\langle \varphi_0\left| V\right| d\rangle =\alpha _{n}\,V_{0d}
\label{12}
\end{equation}
Above, we assume that $\alpha _{n}\equiv \langle d|\varphi _{n}\rangle$
depends exclusively on the energy of the state $\vert\varphi _{n}\rangle$,
namely, $\alpha_{n}\equiv \alpha(\varepsilon _{n})$.

It is convenient to represent the intrinsic Hamiltonian in the space spanned
by the doorway and the subspace orthogonal to it, in the following way:

\begin{equation}
H_{0}=\left| \varphi _{0}\right\rangle \,\varepsilon _{0\,}\,\left\langle
\varphi _{0}\right| \ +\sum\limits_{m}\left| m\right\rangle \Lambda
_m\left\langle m\right| \ +\ \left| d\right\rangle \ \varepsilon
_{d}\,\left\langle d\right| \ +\ \sum\limits_{m}\left\{ \,\left|
m\right\rangle \ \Delta _{m\,}\left\langle d\right| \ +\ \left|
d\right\rangle \,\,\Delta _{m}^{*}\ \left\langle m\right| \right\} \,.
\label{13}
\end{equation}
The last term in the above equation represents the spreading of $\left|
d\right\rangle $. We emphasize that, although $\left\langle m\vert
d\right\rangle=0$, neither $\left| d\right\rangle $ nor $\left|
m\right\rangle $ are eigenstates of the system. We have also ignored
coupling among different $\left| m\right\rangle $ in $H_{0}$.

It is now a simple matter to derive an expression for $\alpha (\varepsilon
_{n})$ and $\Lambda _{m}$. Inserting Eq.~(\ref{13}) into Eq.~(\ref{5}), and
using that $<\varphi _{n^{\prime }}|\varphi _{n}>=\delta _{n^{\prime }n}$,
we obtain

\begin{equation}
\left| \alpha (\varepsilon _{n})\right| ^{2}=\frac{1}{1+\sum_m \frac{\left|
\Delta _{m}\right| ^{2}}{\left( \varepsilon _{n}-\,\Lambda _{m}\right) ^{2}}}
\label{14}
\end{equation}

\noindent and

\begin{equation}
\varepsilon _{n}=\varepsilon _{d}+\sum_{m}\frac{\left| \Delta _{m}\right|
^{2}}{\varepsilon _{n}-\,\Lambda _{m}}  \label{15}
\end{equation}

It is convenient for our purposes to obtain an expression for $\left| \alpha
(\varepsilon _{n})\right| ^{2}$ which contains explicit reference to the
difference $\varepsilon _{n}-\varepsilon _{d}$. For this purpose one must
eliminate $\Lambda _{m}$ from Eqs.~(\ref{14}) and (\ref{15}). This is
analytically possible only for a restricted set of cases. If one uses the
model of ref.~\cite{11}, namely a constant $\Delta _{m}=$ $\Delta $, and a
uniform spectrum [$\Lambda _{m}$] in the sense $\Lambda _{m}=m\,s,$ where $s$
is the constant spacing between adjacent levels and $m=0,\pm 1,\pm 2,...$,
exact expressions for $\varepsilon _{n}$ and $\left| \alpha (\varepsilon
_{n})\right| ^{2}$ can be derived:

\begin{equation}
\varepsilon _{n}=\varepsilon _{d}+\frac{\pi \Delta ^{2}}{s}\,\,\cot \left( 
\frac{\pi \varepsilon _{n}}{s}\right)  \label{16}
\end{equation}
and

\begin{equation}
\mid \alpha (\varepsilon _{n})\mid ^{2}=\frac{1}{1+\left( \pi \Delta
/s\right) ^{2}\csc ^{2}\left( \pi \varepsilon _{n}/s\right) }\,=\,\frac{1}{%
1+\left( \pi \Delta /s\right) ^{2}+\frac{\left( \varepsilon _{d}-\varepsilon
_{n}\right) ^{2}}{\Delta ^{2}}}\,.  \label{17}
\end{equation}
Defining the damping width of the doorway state $\Gamma _{d}^{\downarrow
}=2\pi \Delta ^{2}/s\equiv 2\pi \Delta ^{2}\rho $, where $\rho =1/s$ is the
constant density of the $\vert m\rangle$ states, we may finally write

\begin{equation}
\left| \alpha (\varepsilon _{n}){}\right| ^{2}=\frac{1}{2\pi \rho } \,\,%
\left[ \frac{\Gamma _{d}^{\downarrow }}{\left( \varepsilon _{n}-\varepsilon
_{d}\right) ^{2}+\frac{\Gamma _{d}^{\downarrow 2}}{4} +\Delta ^{2}}\right]
\label{18}
\end{equation}
The term $\Delta ^{2}$ in the denominator is usually absorbed in the
definition of the doorway energy: $\varepsilon _{d}^{\prime }$ $=\varepsilon
_{d}+\Delta ^{2}$. This then results in the celebrated Breit-Wigner formula
for $\left| \alpha (\varepsilon _{n})\right| ^{2},$

\begin{equation}
\left| \alpha \left( \varepsilon _{n}\right) \right| ^{2} =\frac{1}{2\pi
\rho }\,\,\left[ \frac{\Gamma _{d}^{\downarrow }}{\left( \varepsilon
_{n}-\varepsilon _{d}-\Delta \varepsilon _{d}\right) ^{2}+\frac{\Gamma
_{d}^{\downarrow 2}}{4}}\right]  \label{19}
\end{equation}
We may now normalize $\alpha \left( \varepsilon _{n}\right) $,

\begin{equation}
\sum\limits_{n}\left| \alpha \left( \varepsilon _{n}\right) \right|
^{2}\approx \int \mid \alpha \left( \varepsilon _{n}\right) \mid
^{2}\,\,\rho (\varepsilon _{n})\,\,\,\,d\varepsilon _{n}=1  \label{20}
\end{equation}
The normalization above may be easily performed if we make the assumption
that the density of the eigenstates of the system, $\rho (\varepsilon _{n})$%
, equals that of the states orthogonal to the doorway, $\rho =1/s$.

Since the doorway decays into the open channels, one should add to $\Gamma
_{d}^{\downarrow }$ in Eq.~(\ref{18}) the escape width, $\Gamma
_{d}^{\uparrow }$. This can be formally done using reaction theory techniques
\cite{8}. Therefore, the coupling matrix elements will be calculated
according to Eq.~(\ref{12}) with $\left| \alpha (\varepsilon _{n})\right|
^{2}$ given by

\begin{equation}
\left| \alpha (\varepsilon _{n})\right| ^{2}=\frac{1}{2\pi \rho }\,\,\left[ 
\frac{\Gamma _{d}}{\left( \varepsilon _{n}-\varepsilon _{d}-\Delta
\varepsilon _{d}\right) ^{2}+\Gamma _{d}^{2}\,/\,4}\right] \ .  \label{21}
\end{equation}
where 
\begin{equation}
\Gamma _{d}=\Gamma _{d}^{\downarrow }+\Gamma _{d}^{\uparrow }  \label{22}
\end{equation}
is the total width of the doorway state.

The doorway approach described above depends on the incident energy. The
faster the colliding probe is, the less details will be exhibited by the
collision spectrum. If the energy is lowered, one expects that more details
of the cluster structure will be revealed. Namely, the doorway state
described above will fragment, at lower energies, owing to the coupling to
some particular $\vert m\rangle$ states.

The method proposed here should be particularly useful in the collisional
determination of the complicated spectrum of large atomic clusters. This is
because a detailed description of such large many-body system is, in
general, prohibitively complicated.

\section{PLASMON OSCILLATIONS AS DOORWAY CHANNEL FOR COULOMB EXCITATION OF
METALLIC CLUSTERS}

Let us now consider the specific case of an electric dipole resonance. If $%
\varphi _{0\text{ }}$ has angular momentum $j=0$ and the coupling $V$ is not
too strong, only excited states with $j=1$ (in $\hbar $ units) are
significantly populated. In this case, such states can be labeled by their
energies (in principle a continuous label) and the $z$-componente of $j$, $%
\mu =0,\pm 1$; namely: $n\equiv \varepsilon ,\mu .$ Thus, we get the coupled
channels equations

\begin{equation}
\frac{d a_0(t)}{dt}=\frac{1}{i\hbar }\sum_\mu \int d\varepsilon
\,\,\,e^{-\,i\varepsilon \,t\,/\,\hbar }\,\,V_{0,\varepsilon \mu
\,\,\,\,}a_{\varepsilon \mu }(t)  \label{23}
\end{equation}

\begin{equation}
\frac{da_{\varepsilon \mu }(t)}{dt}=\frac{1}{i\hbar }\,\,\,e^{i\varepsilon
\,t\,/\,\hbar \,\,}V_{\varepsilon \mu ,0\,\,}a_{0}\left( t\right) \,,
\label{24}
\end{equation}
with the initial condition $a_{n}(t\rightarrow -\infty )=\delta _{n,0}$.
Above, we have used the fact that the dipole coupling does not mix the
excited states. This is so because the dipole operator is a spherical tensor
of rank 1 and all these states have $j=1$. In this case the excitation
occurs through the three degenerate doorways $\left| d_{\mu =0,\pm
1}\right\rangle$, with energy $\varepsilon_d$ and angular momentum $j=1$.
These doorways differ only in the $z$-component of the angular momentum, $%
\mu=0,\pm 1$. We assume that the expansion coefficients $\alpha (\varepsilon
)$ are the same for the three doorways, independently of $\mu $. The
coupling matrix elements are then written 
\begin{equation}
V_{\varepsilon \mu ,0\,\,}=\alpha (\varepsilon )\,\left\langle
d_{\mu}\right| V(t)\left| \varphi_0\right\rangle \equiv \alpha
(\varepsilon)\,\,V_{\mu }(t)\,.  \label{25}
\end{equation}

Integrating Eq.~(\ref{24}) over time, substituting the result in Eq.~(\ref
{23}) and using Eq.~(\ref{25}), we obtain

\begin{equation}
\frac{da_{0}\left( t\right) }{dt}=\frac{1}{\hbar ^{2}}\,\,\sum_{\mu
}\,V_{\mu }(t)\,\,\int d\varepsilon \ \left| \alpha \left( \varepsilon
\right) \right| ^{2}\,\int_{-\infty }^{t}dt^{\prime
}\,\,\,\,e^{-\,i\,\varepsilon \,(t-t^{\prime })/\,\hbar }\,\,\left[ V_{\mu
}(t^{\prime })\right] ^{*}\,\,\,a_{0}(t^{\prime })\,.  \label{26}
\end{equation}
Assuming the Breit-Wigner shape for $\left| \alpha (\varepsilon )\,\right|
^{2},$ the integration over $\varepsilon $ can be carried out in the complex
plane. We find,

\begin{equation}
\frac{da_{0}\left( t\right) }{dt}=\frac{1}{\hbar ^{2}}\,\sum_{\mu
}V_{\mu}(t)\, \int_{-\infty }^{t}dt^{\prime }\,\,\,\,e^{-\,i\,\left(
\varepsilon _{d}-\, i\,\Gamma _{d}/2\right) \,(t-t^{\prime })\,/\,\hbar }\,\,%
\left[ V_{\mu } (t^{\prime })\right] ^{*}\,\,\,a_{0}(t^{\prime })\, .
\label{27}
\end{equation}
This equation can be solved by introducing the auxiliary amplitudes $A_{\mu
}\left( t\right) $, such that $\,\,a_{0}\left( t\right) =1+\sum_{\mu }A_{\mu
}\left( t\right) $. Eq.~(\ref{27}) then reduces to the set of three coupled
differential equations 
\begin{equation}
\ddot{A}_{\mu }\left( t\right) +\left[ \frac{\stackrel{.}{V}_{\mu }\left(
t\right) }{V_{\mu }\left( t\right) }-\frac{i}{\hbar }\left( \varepsilon _{d}-%
\frac{i}{2}\Gamma _{d}\right) \right] \,{\dot{A}}_{\mu }\left( t\right) +%
\frac{\mid V_{\mu }\left( t\right) \mid ^{2}}{\hbar ^{2}}\left( 1+\sum_{\mu
}A_{\mu }\left( t\right) \right) =0\,,  \label{28}
\end{equation}
with initial conditions $A_{\mu }\left( t\rightarrow -\infty \right) =0$.
Eq.~(\ref{28}) can be handled by standard numerical procedures. Once
theasymptotic amplitude $a_{0}\left( t\rightarrow \infty \right)$ is
calculated, the excitation probabilities are easily obtained from Eqs.~(\ref
{10}) and (\ref{24}) (integrated over time)

\begin{equation}
P_{\varepsilon ,\mu }\equiv \left| a_{\varepsilon \mu }\left( t\rightarrow
\infty \right) \right| ^{2}=\frac{1}{2\pi }\,\;\left[ \,\frac{\Gamma _{d}}{%
\left( \varepsilon _{n}-\varepsilon _{d}\right) ^{2}+\frac{\Gamma _{d}^{2}}{4%
}}\right] \left| \int_{-\infty }^{\infty }dt\,e^{\,i\varepsilon
\,\,t\,/\,\hbar }\,\,\left[ V_{\mu }(t)\right] ^{*}\,\,\,a_{0}(t)\right|
^{2}\,.  \label{29}
\end{equation}
Integrating over the impact parameter (implicit in the dependence of $V(t)$
on the trajectory), we obtain the excitation cross sections

\begin{equation}
\frac{d\sigma _{\mu }}{d\varepsilon }=2\pi \,\sum_{\mu
}\int_{b_{min}}b\,db\,P_{\varepsilon ,\mu }(b)  \label{30}
\end{equation}
In Eq. (\ref{30}), $\sigma _{\mu }$ stands for the cross section for
population of states with quantum number $\mu $, and $b_{min}$ represents
the value of $b$ below which trajectories suffer strong inelastic
transitions. We take $b_{min}$ to be an effective radius of the cluster core 
$\approx l\ {\cal N}^{1/3}$ where $l$ is the mean interatomic distance in
the cluster and ${\cal N}$ is the number of atoms.

\section{APPLICATIONS}

Now we apply the formalism of the previous sections to the plasmon resonance
in Sodium clusters, identified here as the doorway state. The resonance is
excited in collisions with electron and proton projectiles. We consider the
excitation in both neutral and ionized clusters. In the former case, one can
use straight line trajectories for ${\bf R}(t)$ while in the latter one
should use Coulomb trajectories.

First we write the explicit expression for the coupling, $V_{\mu }$. Using
the dipole approximation and writing Eq.~(\ref{2}) in terms of the spherical
components of the dipole field $E_\mu (t)$ and of the dipole operator $D_\mu$%
, we get

\begin{equation}
V = - \sum_\mu\ E_\mu (t)\ \cdot \ D_\mu\, ,
\end{equation}
with

\begin{equation}
E_\mu (t) = E(t) \ \sqrt{\frac{4\pi}{3}}\ Y^*_{1\mu}({\hat {{\bf R}}}(t))\ ,
\end{equation}

where $E(t)$ is the electric field strength 
\begin{equation}
E(t) = \frac{Ze^2}{R^2\left( t\right) }\, .  \label{33}
\end{equation}

If we use the reference frame of figure 1, where the z-axis if orthogonal to
the collision plane, the x-axis is along the apex-line towards the
projectile and the initial velocity has positive y-component, the matrix
element $V_{\mu =0}$ vanishes identically\cite{AW}. Using Wigner-Eckart's
theorem and representing the other two components with the shorthand
notation $V_{\mu =\pm 1}\equiv V_{\pm }$, we obtain

\begin{equation}
V_{+}=\frac{\mid \langle \varphi_0\mid\mid D\mid\mid d\rangle \mid}{\sqrt{2}}
\,\ E(t)\ \, e^{-i\phi \left( t\right) }\, ,  \label{34}
\end{equation}

\begin{equation}
V_{-}=V_{+1}^*\, .  \label{35}
\end{equation}
The phase $\phi (t)$ is given by

\begin{equation}
\phi (t)=- \frac{\pi-\Theta}{2}\ + \ bv\int_{-\infty }^t\frac{dt^{\prime }}{%
R^2\left( t^{\prime } \right)}\, ,  \label{36}
\end{equation}
where $Z_c$ is the charge of the projectile and $\Theta$ is the scattering
angle.

The reduced matrix-element $\langle \varphi_0\mid\mid D\mid\mid d\rangle $
can be estimated from the energy weighted sum rule, applied to collective
oscillations in the small sodium clusters\cite{4}-\cite{6}. It leads to

\begin{equation}
\langle \varphi_0\mid\mid D\mid\mid d\rangle = e\,\ \sqrt{\frac{\hbar^2 
{\cal N} K}{2m\, \varepsilon_d }}\;,
\end{equation}
where $K$ is the fraction of the total oscillator strength exhausted by the
plasmon mode, ${\cal N}$ is the number of atoms in the clusters, and $m$ is
the electron mass. As an illustration, we consider a small Sodium cluster
with ${\cal N}=8$ atoms. In this case, we have the typical values: $%
\varepsilon_d\approx 2.5\ {\rm eV} =0.093\ {\rm a.u.}$, $\Gamma_d=0.1\
\varepsilon_d $, $K\approx 0.7$, and the interatomic distance $l\approx 3$
\AA . Since we are dealing with a peripheral process, we use as the lower
limit for the impact parameter integration of Eq.~(\ref{30}) the radius of
the cluster. In the present case, $b_{min}=R_{cluster}\approx 7.5$~a.u..

\subsection{Excitation of a neutral atomic metal cluster}

Let us consider the excitation of the plasmon resonance in a metal cluster
with net charge equal to zero. In this case it is reasonable to use a
straight line trajectory for the projectile motion. In the reference frame
of figure 1, the trajectory reduces to a straight line parallel to the
y-axis, so that

\begin{equation}
{\bf R}(t)=b\ {\bf \hat{x}}\ +\ vt \ {\bf \hat{y}}\,.
\end{equation}
The electric field strength (Eq.~(\ref{33})) and the phase $\phi (t)$ (Eq.~(%
\ref{36})) are then given by

\begin{equation}
E(t) = \frac{Ze^2}{b^2\ +\ v^2\ t^2}  \label{39}
\end{equation}
and

\begin{equation}
\phi (t) = \arctan\left(\frac{vt}{b}\right)\, .  \label{40}
\end{equation}

We have solved numerically Eq.~(\ref{28}) taking as projectiles electrons
with incident energy $E=100$~eV and protons with $E=100$~keV. In figures 2a
and 2b, we show the time evolution of the probabilities for the cluster to
remain in the ground state, $P_0$, and to be excited to a state with each of
the possible values of $\mu$, $P_1$ and $P_{-1}$. The latter where obtained
by summation over the excitation energy. The results where obtained for the
impact parameter $b=7.5$ a.u. It is interesting to notice that the
population probability for $\mu =-1$ is much larger than that for $\mu =1$.

In figures 3a and 3b, we give the calculated excitation spectra for the same
projectiles and energies. We notice that the results are rather similar,
despite of the different sign of the projectile charge in each case. The
reason is that the cross section depends basically on the projectile
velocity, which is of the same order of magnitude in the two cases. We have
found that effect of the width $\Gamma _{d}$ (taken here as $\Gamma
_{d}=0.1\ \hbar \omega _{d}$ \cite{2,4,7}) on the cross sections is very
small. For all practical purposes one may set $\Gamma _{d}=0$ in the coupled
channel calculation, using the doorways as eigenstates of $H_{0}$, and then
use Eq.~(\ref{29}) for the probabilities $P_{\varepsilon \mu }$.

Consistently with the probabilities of figures 2a and 2b, we find that the
contribution of the state with $\mu =-1$ completely dominates the cross
section. This can be understood if one neglects the width $\Gamma _{d}$ and
use first order perturbation theory to calculate the excitation amplitudes.
In this case the amplitudes $a_{\pm }(t)\equiv a_{\mu =\pm 1}$ are obtained
by integrating Eq.~(\ref{8}) with $a_{0}(t)=1$. Using the explicit form of
the matrix elements (Eqs.~(\ref{33}) and (\ref{34})), we get

\begin{equation}
a_{\pm}(t)=-\frac{i}{\hbar}\ \ \frac{<\varphi_0\ \vert\vert\ D\ \vert\vert\
d>} {\sqrt{2}}\ \int_{-\infty}^t\ dt^{\prime}\ E(t^{\prime})\ \exp\left[ 
\frac{i}{\hbar} \left( \varepsilon_d t^{\prime}\pm \hbar
\phi(t^{\prime})\right)\right]\, .  \label{41}
\end{equation}
Using Eq.~(\ref{39}) and (\ref{40}) we find that the stationary phase points
of the integrand are given by the condition

\begin{equation}
\varepsilon_d=\mp\ \frac{\hbar v}{b}\, \left[\frac{b^2}{b^2+v^2t_\pm^2} %
\right] \, ,
\end{equation}
where $t_\pm$ stands for the stationary point in the integrand giving the
amplitude $a_\pm$. Since the energy $\varepsilon_d$ is positive, $a_+$ has
no stationary point. For $a_-$, the stationary point is

\begin{equation}
t_{-}=\sqrt{\frac{\hbar b}{v\varepsilon _{d}}-\frac{b^{2}}{v^{2}}}\,.
\end{equation}
Within the stationary phase approximation, the amplitude $a_{+}$ vanishes,
consistently with the low probability and cross section for $\mu =+1$ in
figures 2a, 2b, 3a and 3b.

It is important to point out that the valitity of the stationary phase
approximation requires that the phase in Eq.~(\ref{41}) varies rapidly
within the collision time, $\tau _{c}$. The opposite situation would be the
sudden limit where the doorway energy is negligible, i.e. $\varepsilon
_{d}<<\hbar /\tau _{c}$. In this case, it can be easily seen that $%
a_{-}\simeq a_{+}$ and $\sigma _{_{-}}\simeq \sigma _{_{+}}$.

\subsection{Excitation of a charged metal cluster}

We now consider the excitation of the plasmon mode in a ionized metal
cluster with net charge $Z_{c}e,$ in collisions with the same projectiles of
the previous section. All the parameters of the metal cluster are the same
as in the previous subsection. In this case, ${\bf R}(t)$ is a Rutherford
trajectory. We first consider the case where the Coulomb field is repulsive (%
$Z$ and $Z_{c}$ have the same sign). We introduce the Sommerfeld parameter

\begin{equation}
\eta =\frac{Z\,Z_{c}\,e^{2}}{\hbar v},  \label{43}
\end{equation}
the half-distance of closest approach in a head-on collision,

\begin{equation}
a=\frac{Z\,Z_{c}e^{2}}{2\,E}  \label{44}
\end{equation}
and the excentricity,

\begin{equation}
\epsilon =\sqrt{1+\frac{b^{2}}{a^{2}}}.
\end{equation}
It is conveniente to introduce a new variable $w$, defined by its relation
with the time,

\begin{equation}
t=\frac{a}{v}\,\left[ \epsilon \sinh w+w\right] \,.  \label{47}
\end{equation}
>From the equations of motion under the action of the Coulomb field, it is
easy to show that the trajectory on the $xy$-plane, specified by the
coordinates $R$ and $\phi $ (see figure 1), is given in terms of $w$ as

\begin{eqnarray}
R(w)&=&a\,\left[ \epsilon \,\cosh w+1\right]  \nonumber \\
\phi (w)&=&\arcsin \left( \frac{a\,\sqrt{1-\epsilon ^{2}}\,\,\sinh w\,}{R(w)}
\right) \,.  \label{48}
\end{eqnarray}

Using the variable $w$ in Eq.~(\ref{28}) we get

\begin{equation}
A_{\mu }^{\prime \prime }\left( w\right) +\frac{1}{f(w)}\left[ \frac{%
\stackrel{.}{V_{\mu }}\left( t{\scriptsize (w)}\right) }{V_{\mu }\left( t%
{\scriptsize (w)}\right) }-\frac{i}{\hbar }\left( \varepsilon _{d}-\frac{i}{2%
}\Gamma _{d}\right) \right] A_{\mu }^{\prime }(w)+\left| \frac{V_{\mu
}\left( t{\scriptsize (w)}\right) }{\hbar f(w)}\right| ^{2}\left(
1+\sum_{\mu }A_{\mu }\left( w\right) \right) =0\,,
\end{equation}
where primes stand for derivation with respect to $w$ and

\begin{equation}
f(w)=\frac{dw}{dt}=\frac{v}{R(w)}\,.
\end{equation}

>From the amplitudes $A_{\mu }\left( w\right) $, we obtain 
\begin{equation}
a_{0}\left( w\right) =1+\sum_{\mu }A_{\mu }\left( w\right)  \label{51}
\end{equation}
and

\begin{equation}
a_{\varepsilon \mu }(w)=\frac{1}{i\hbar }\;\left[ \alpha (\varepsilon)\right]
^{*}\; \int_{-\infty }^{w}e^{i\varepsilon \,\,t(w^{\prime })\,/\,\hbar \,\,}%
\left[ V_{\mu\,} (w^{\prime })\right] ^{*}\,\,a_{0}\left( w^{\prime }\right)
\;\frac{dw^{\prime }} {f(w^{\prime})}\,.  \label{52}
\end{equation}

For attractive Coulomb interactions, it is necessary to make the changes: 
\begin{eqnarray}
t&=&\frac{a}{v}\,\left[ \epsilon \sinh w-w\right]  \nonumber \\
R(w)&=&a\,\left[ \epsilon \,\cosh w-1\right] \, ,  \label{53}
\end{eqnarray}
the other equations remaining the same as for the repulsive Coulomb field.

We have performed calculations for the excitation of the plasmon mode in the
same cluster and for the same projectiles as in the previous sub-section.
However, here we consider an ionized Sodium cluster. In this case, the
projectile's trajectory is no longer a straight line but a hyperbola. To
assess the importance of this effect we have calculated the total plasmon
excitation cross sections (integrated over excitation energy and summed over 
$\mu $) as a function of the collision energy and of the cluster's net
charge. In this simple example, we have disregarded the effects that the
cluster charge might have on its structure.

The results are shown in figures 4a, for electrons, and in 4b, for protons.
In collisions with electrons, we notice that the cross section decreases
with the collision energy in the energy range considered. Although the
deviations from results using a straight line trajectory (curve with $Z_{c}=0
$) are very small for high energies, Coulomb effects on the trajectory play
an increasingly important role as the collision energy decreases. The
situation for proton projectiles is somewhat different for the collision
energies considered in figure 4b. First, we notice that the cross section
reaches a maximum at $E\approx 25$~keV. The second difference is that the
cross section is not affected by the cluster's net charge. In this case the
curves for 0, 1, ...,8 units of charge cannot be distinguished in the
figure. This result is not surprising. If we estimate the maximal deflection
angle $\Theta $ involved in the graph: for maximal charge ( $Z_{c}=8),$
minimal impact parameter ($b=b_{min}=7.5$ a.u.) and minimal energy ($E=20$
keV), we find $\Theta _{\max }=0.04\ \deg .$ Therefore, the Coulomb
trajectory is very closely a straight line.

\section{CONCLUSIONS}

We have developed a general theoretical framework through which the
collisional excitations of collective plasmon resonances in metallic
clusters can be quantitatively described. The idea of a doorway state
representing the resonance, which is coupled to both the open decay channels
and the chaotic degrees of freedom of the cluster, is used for this purpose.
Application of the semiclassical scattering theory in conjunction with the
``exit'' doorway model is made to light Na-clusters using electrons and
protons as projectiles. Although Pauli exchange effects in the electron
scattering are important and they result in non-local potentials, we have
ignored these effects here for simplicity. The formalism can be applied
easily to cluster-cluster collisions. Furthermore, it can be generalized to
cases involving the excitation of multiplasmon states in both spherical and
deformed clusters. In the deformed cluster case one sees a two-peaked one
plasmon resonance\cite{1} and should see a three-peaked two-plasmon
resonance, a four-peaked three-plasmon resonance etc.\cite{11}. These
resonances can be reached in multistep processes (ground state$\rightarrow $
one plasmon$\rightarrow $ two plasmon$\rightarrow $ etc.).

In our theory only the average cross-section corresponding to the excitation
of the doorway state was calculated. There should be fluctuation
contributions, not considered here. Such contributions will be discussed in
future work.

\vskip 1cm

\noindent This work was supported in part by CNPq, FAPESP, the National
Science Foundation, the PRP/USP and the MCT/FINEP/CNPq(PRONEX) under
contract no. 41.96.0886.00.

\vfill
\eject

\centerline {\bf Figure Captions.}

\begin{itemize}
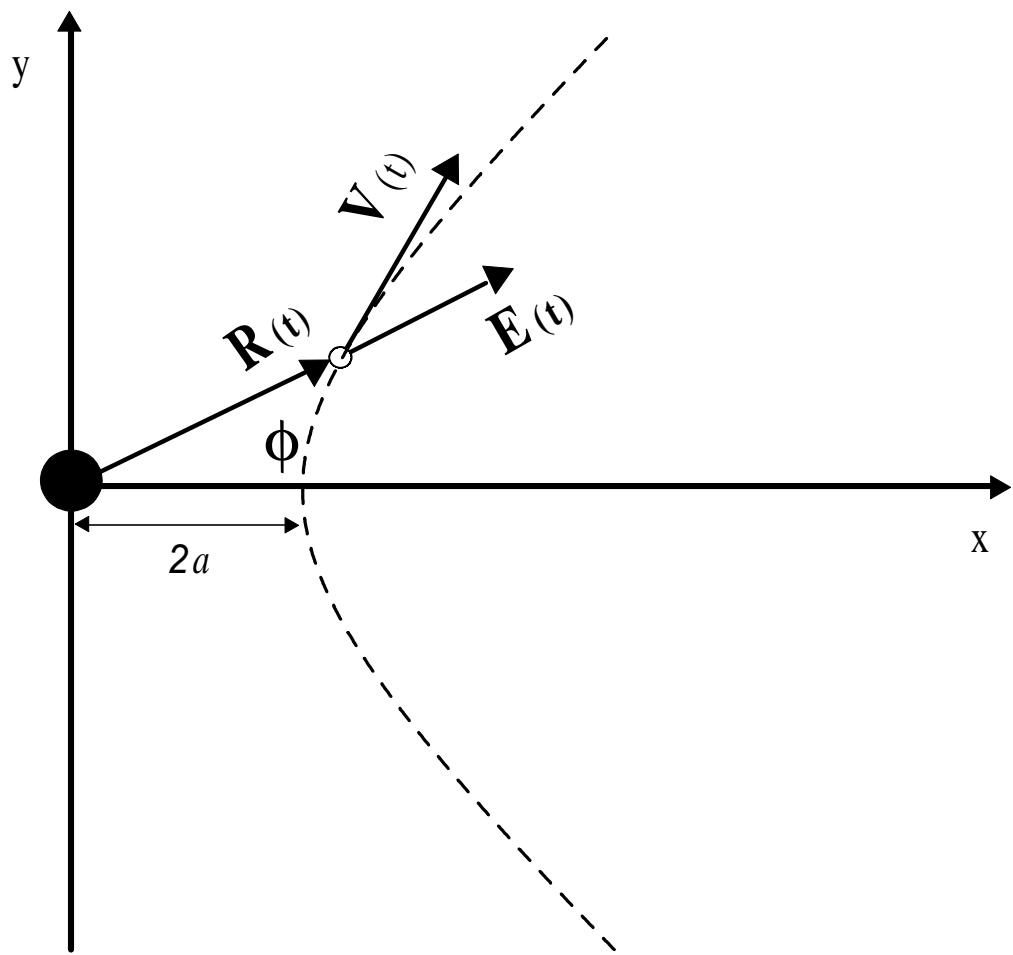
\item  {Figure 1:} Collision of a charged projectile (open circle) with
metal cluster (solid circle). In this illustration, the charge of the
cluster has the same sign as that of the projectile. In this reference
frame, the trajectory lies on the {\it xy}-plane and the {\it z}-axis points
upwards, away from the paper sheet. The coordinates {\it R} and $\phi $, and
the distance of closest approach, $2a$ are indicated.

\item  {Figure 2a:} Time evolution of the elastic and inelastic
probabilities in collisions of 100 eV electrons with a Na-cluster. The time
is indicated in atomic units (a.u.), defined as the ratio of the radius of
the ground state Bohr orbit in the Hydrogen atom to the electron speed in
this state.

\item  {Figure 2b:} Same as figure 2a for 100 keV protons.

\item  {Figure 3a:} The differential cross section as a function of the
cluster excitation energy in the plasmon region. The projectile is a 100 eV
electron. For details see the text. The cross section is given in atomic
units, which in this case are defined as the radius square of the ground
state Bohr orbit in Hydrogen.

\item  {Figure 3b:} Same as figure 3a for 100 keV protons.

\item  {Figure 4a:} Cross sections for the excitation of the plasmon
resonance in ionized Sodium clusters. The results are given as functions of
the collision energy, for a few different values of the cluster's net
charge. For details see the text.

\item  {Figure 4b:} Same as Figure 4a for protons. Here the curves for
different cluster charges cannot be distinguished.
\end{itemize}

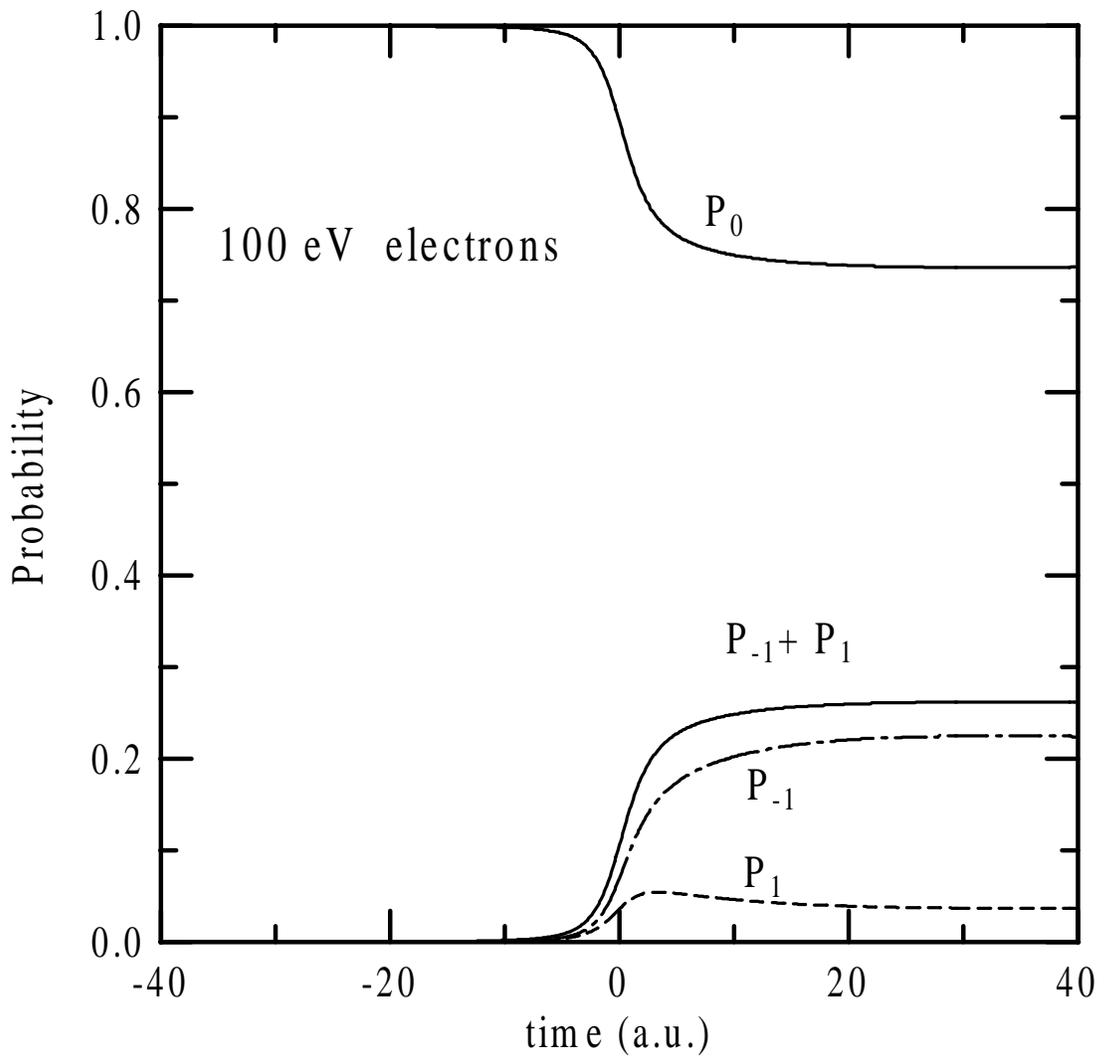
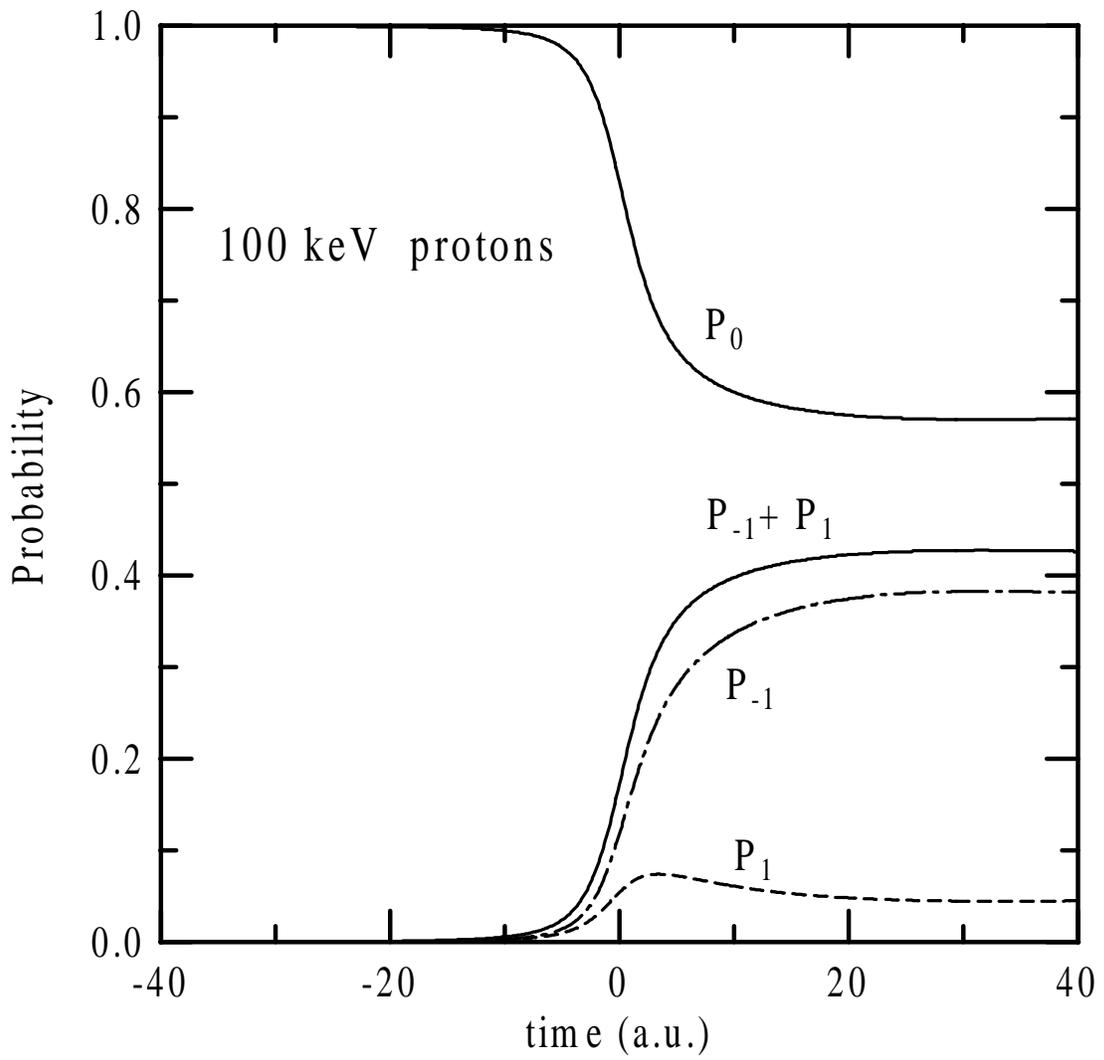
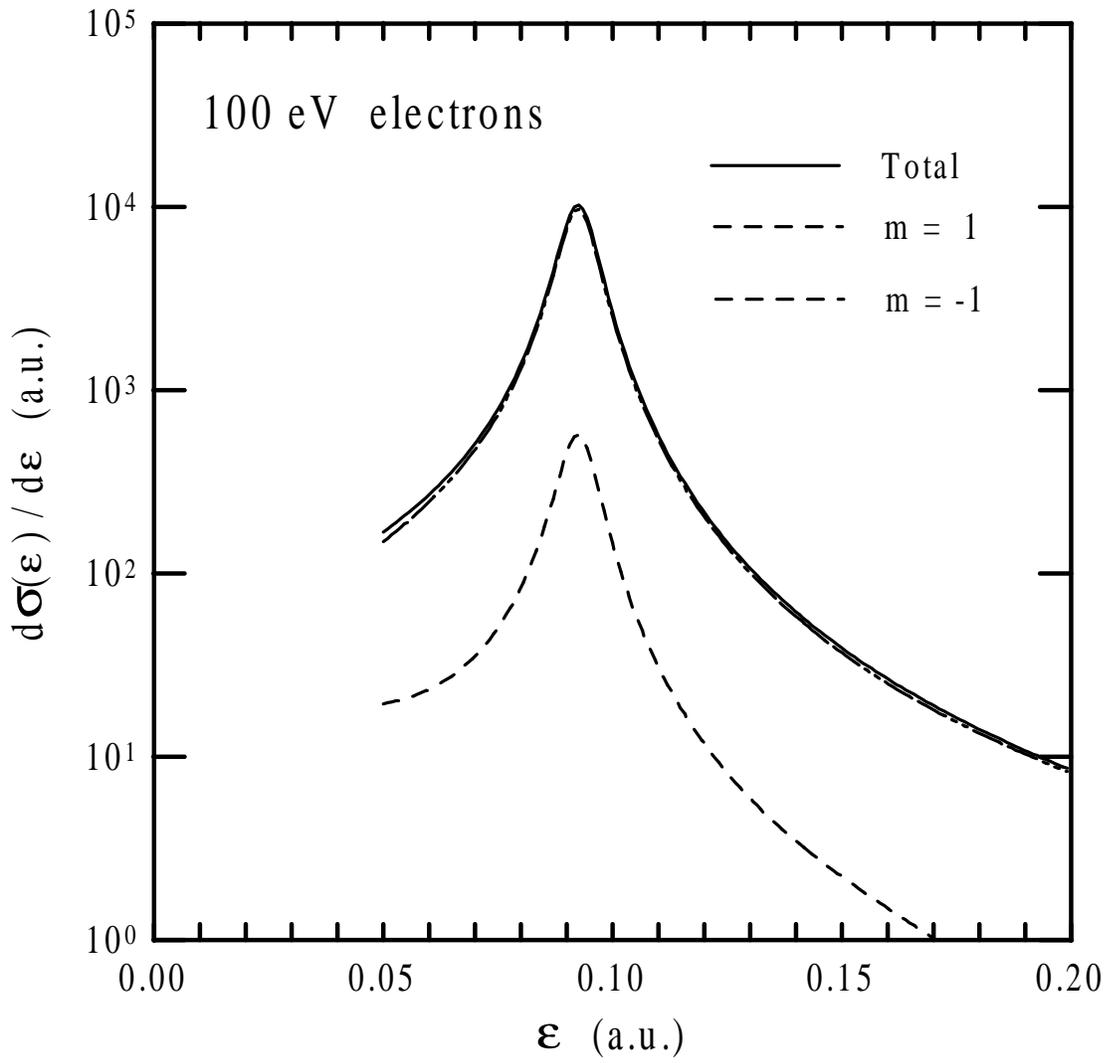
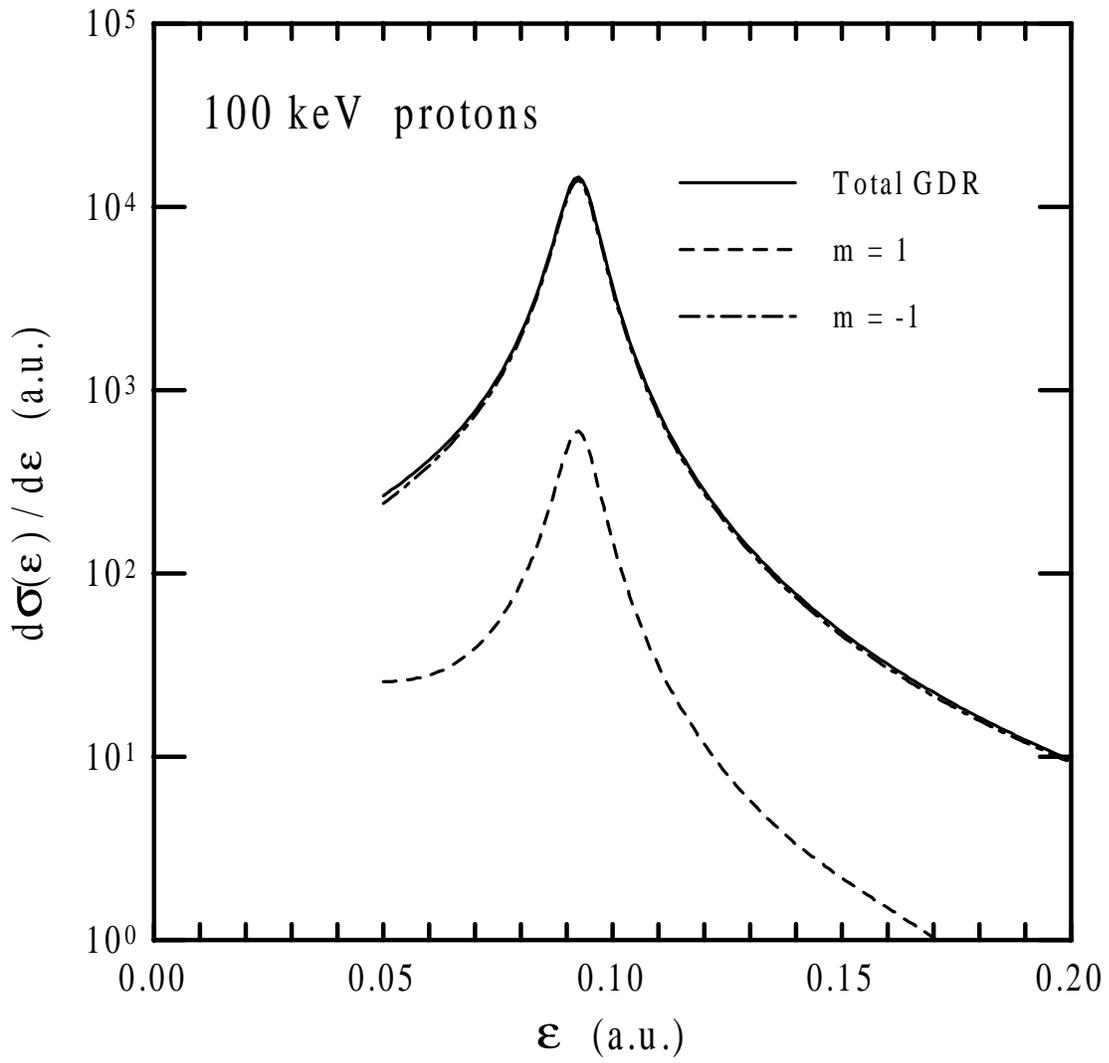
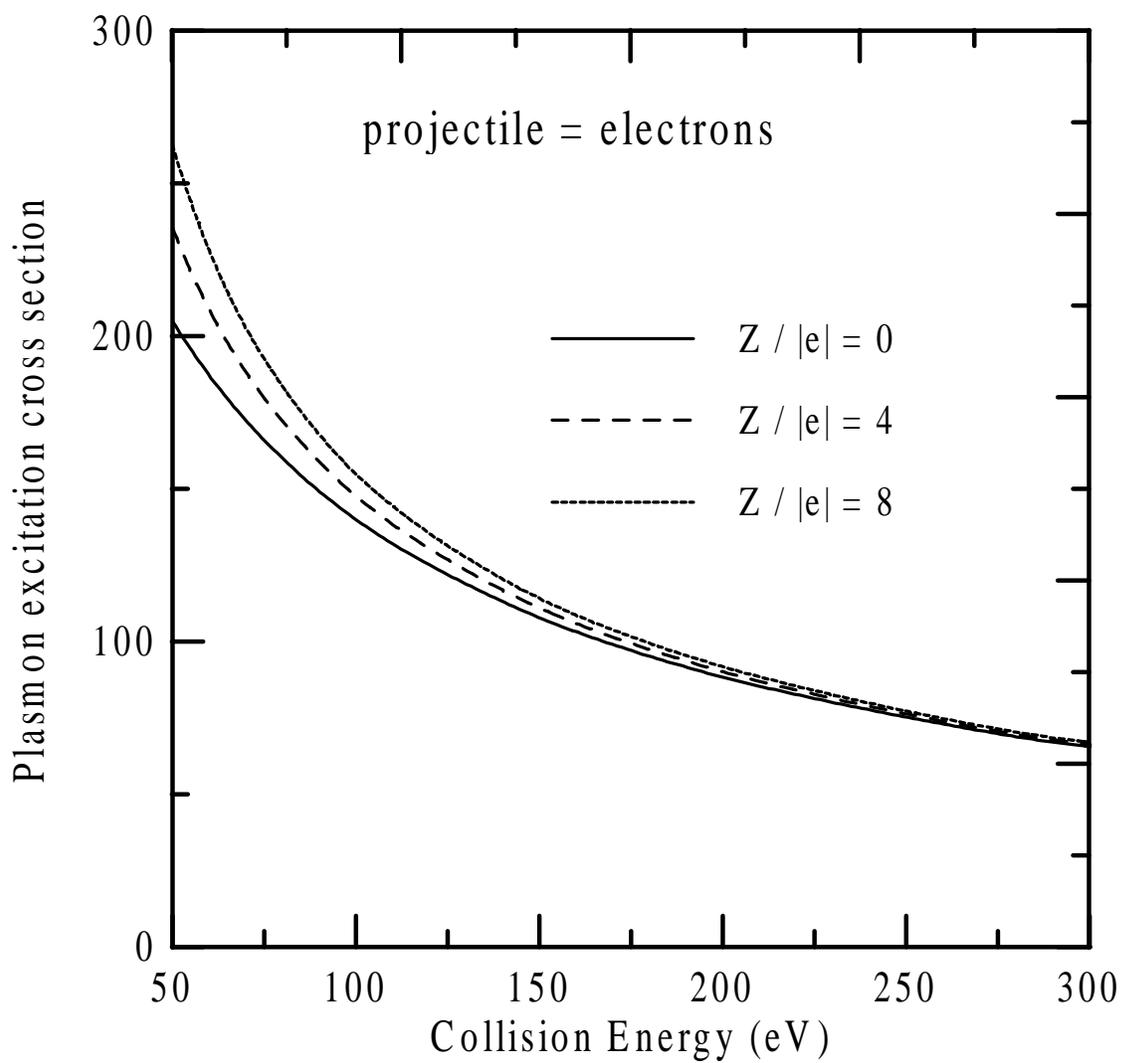
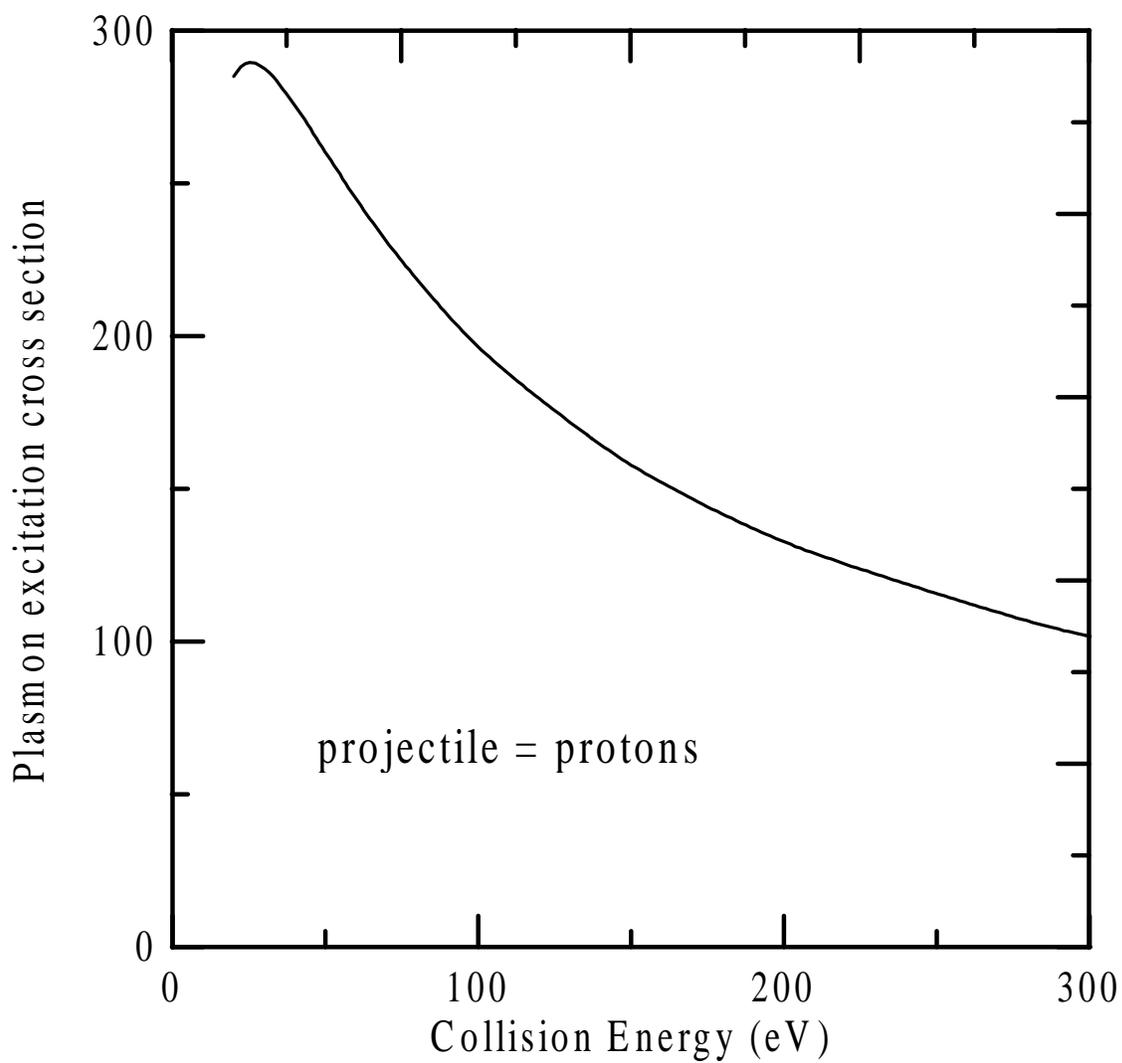
\end{document}